\documentclass[journal]{IEEEtran}
\usepackage{hyperref}       
\usepackage{url}            
\usepackage{booktabs}       
\usepackage{amsfonts}       
\usepackage{nicefrac}       
\usepackage{microtype}      

\usepackage{cite}
\usepackage{amsmath,amssymb,amsfonts} 
\usepackage{algorithmic}
\usepackage{array}
\usepackage{graphicx}
\usepackage{textcomp}
\usepackage{xcolor,colortbl}
\usepackage{multirow}
\usepackage{tabularx}
\usepackage{float}
\usepackage{caption}
\usepackage{subcaption}
\usepackage{cleveref}
\usepackage{comment}
\definecolor{greenC}{rgb}{0.7,1,0.7}

\begin{document}

\title{Robust Imaging of Speed-of-Sound\\ Using Virtual Source Transmission
}%

\author{
Dieter Schweizer,
Richard Rau, Can Deniz Bezek, Rahel A. Kubik-Huch,
Orcun Goksel

\thanks{D.~Schweizer and O.~Goksel are with the Computer-assisted Applications in Medicine Group, ETH Zurich, Switzerland.}%
\thanks{C.~Bezek and O.~Goksel are with the Department of Information Technology, Uppsala University, Sweden.}%
\thanks{R.~A.~Kubik-Huch is with the Kantonsspital Baden, Switzerland.}%
\thanks{R.~Rau contributed the work while he was with the Computer-assisted Applications in Medicine Group, ETH Zurich, Switzerland.}
\thanks{Corresponding author: D.~Schweizer (dieter.schweizer@vision.ee.ethz.ch)}
}

\maketitle

\begin{abstract}
Speed-of-sound (SoS) is a novel imaging biomarker for assessing biomechanical characteristics of soft tissues.
SoS imaging in pulse-echo mode using conventional ultrasound systems with hand-held transducers has the potential to enable new clinical uses.
Recent work demonstrated diverging waves from single-element (SE) transmits to outperform plane-wave sequences. 
However, single-element transmits have severely limited power and hence produce low signal-to-noise ratio (SNR) in echo data.
We herein propose Walsh-Hadamard (WH) coded and virtual-source (VS) transmit sequences for improved SNR in SoS imaging.
We additionally present an iterative method of estimating beamforming SoS in the medium, which otherwise confound SoS reconstructions due to beamforming inaccuracies in the images used for reconstruction.
Through numerical simulations, phantom experiments, and in-vivo imaging data, we show that WH is not robust against motion, which is often unavoidable in clinical imaging scenarios.
Our proposed virtual-source sequence is shown to provide the highest SoS reconstruction performance, especially robust to motion-artifacts.
In phantom experiments, despite having a comparable SoS root-mean-square-error (RMSE) of 17.5 to 18.0\,m/s at rest, with a minor axial probe motion of $\approx$0.67\,mm/s the RMSE for SE, WH, and VS already deteriorate to 20.2, 105.4, 19.0\,m/s, respectively; showing that WH produces unacceptable results, not robust to motion.
In the clinical data, the high SNR and motion-resilience of VS sequence is seen to yield superior contrast compared to SE and WH sequences.
\end{abstract}


\section{Introduction}
\IEEEPARstart{T}{issue} differentiation by ultrasound (US) imaging is indispensable in clinics, both for diagnosis and image-guided interventions. 
B-mode imaging is typically used for tissue differentiation, but this is inherently not a quantitative imaging technique and in some cases does not provide the desired diagnostic information.
For biomechanical tissue characterization, elastography techniques have been increasingly used, nevertheless shear-modulus that these methods typically quantify is not always correlated with pathologically changes, e.g., in breast cancer and liver steatosis.
Speed-of-sound (SoS) in tissues can be used as an alternative quantitative biomarker~\cite{Bamber_ultrasonic_79} and it has been shown to provide better tissue characterization~\cite{glozman_method_2010}, e.g., with higher specificity in breast tumor differentiation~\cite{li_vivo_2009}.

Estimation of a global SoS value in the imaged tissue or layer-wise approximations thereof have been studied by several researchers, especially in the context of correcting aberrations due to SoS~\cite{Ophir_Estimation_86,Anderson_direct_98,Krucker_sound_04,Byram_method_12}.
Several global SoS estimation methods employ trial-and-error to optimize some quality-metric mostly based on a form of speckle analysis~\cite{Napolitano_sound_06,Shin_estimation_10,Qu_average_12,Shen_ultrasound_20} or phase variance~\cite{Yoon_in-vitro_11,Perrot_DAS_21}, although model-based methods have also been proposed~\cite{Xenia_Estimating_21,Bezek_global_22}.
In contrast, for reconstructing maps of local SoS variation, different mechanical setups have been proposed including transmission-mode~\cite{goss_comprehensive_1978, pratt_sound-speed_2007, duric_detection_2007,gemmeke_3d_2007,malik_quantitative_2018,malik_breast_2019},  hand-held reflector-based~\cite{sanabria_hand-held_2016}, and reflector-free pulse-echo mode~\cite{jaeger_computed_2015,sanabria_spatial_2018} systems and methods. 
The latter line of work does not require hardware that is additional to a standard ultrasound system, and hence is ideal for integration into existing clinical workflows, e.g. as a different imaging mode and a potential add-on to standard B-mode, Doppler, and elastography imaging.

Pulse-echo mode SoS imaging methods observe a tissue region of interest from two or more different points of views, e.g. via different transmission (Tx) sequences. 
Any SoS heterogeneities along the US beam paths then cause misalignments in these images.
Having identified these misalignments, one can use a frequency-domain \cite{jaeger_computed_2015} or a spatial-domain \cite{sanabria_spatial_2018} reconstruction method to estimate the spatial SoS distribution (map).
Plane-wave~\cite{montaldo_coherent_2009} and diverging-wave~\cite{jensen_synthetic_2006} Tx both cover large areas of interest and therefore allow for observing the tissue with few numbers of Tx cycles, making them good candidates for SoS imaging.
In~\cite{Rau_divergingWave_2021}, single-element based diverging wave (DW) transmission was proposed for SoS imaging, while showing this to produce less aberration artifacts compared to plane-wave transmission, hence resulting in better SoS reconstructions. 

Most earlier works demonstrate SoS reconstructions with mechanically-fixed workbench experiments.
Nevertheless, such experimental results do not always translate to hand-held SoS imaging nor to in-vivo results.
Indeed, compared to the experimental results of~\cite{sanabria_spatial_2018} with a test-bench fixed setup, a substantial reduction in quality is observed with its translation for in-vivo use in~\cite{Ruby_breast_19}.
From our experience and as also demonstrated later in the results herein, motion and tremor during hand-held ultrasound examination substantially deteriorate SoS imaging.
Therefore, for a successful clinical translation, imaging techniques and sequences insensitive or robust to motion are necessitated.

In this work, we show that single-element (SE) Tx~\cite{Rau_divergingWave_2021}, due to its limited power capability and hence low signal-to-noise ratio (SNR) in the echo data, result in suboptimal imaging of SoS, in particular for complex tissue structures intrinsic in-vivo. 
We herein present two approaches to remedy this and improve SoS reconstruction SNR: a Walsh-Hadamard (WH) coded and a virtual-source (VS) transmit sequence.
We study these sequences in realistic imaging conditions by comparing them to SE Tx, and show that the VS sequence provide the most accurate SoS reconstructions, in particular in presence of motion, which is unavoidable during ultrasound imaging with a hand-held probe in the clinical setting.

\section{Methods}

\subsection{Pulse-echo imaging of speed-of-sound}
\begin{figure*}
  \centering
    \includegraphics[width=\textwidth]{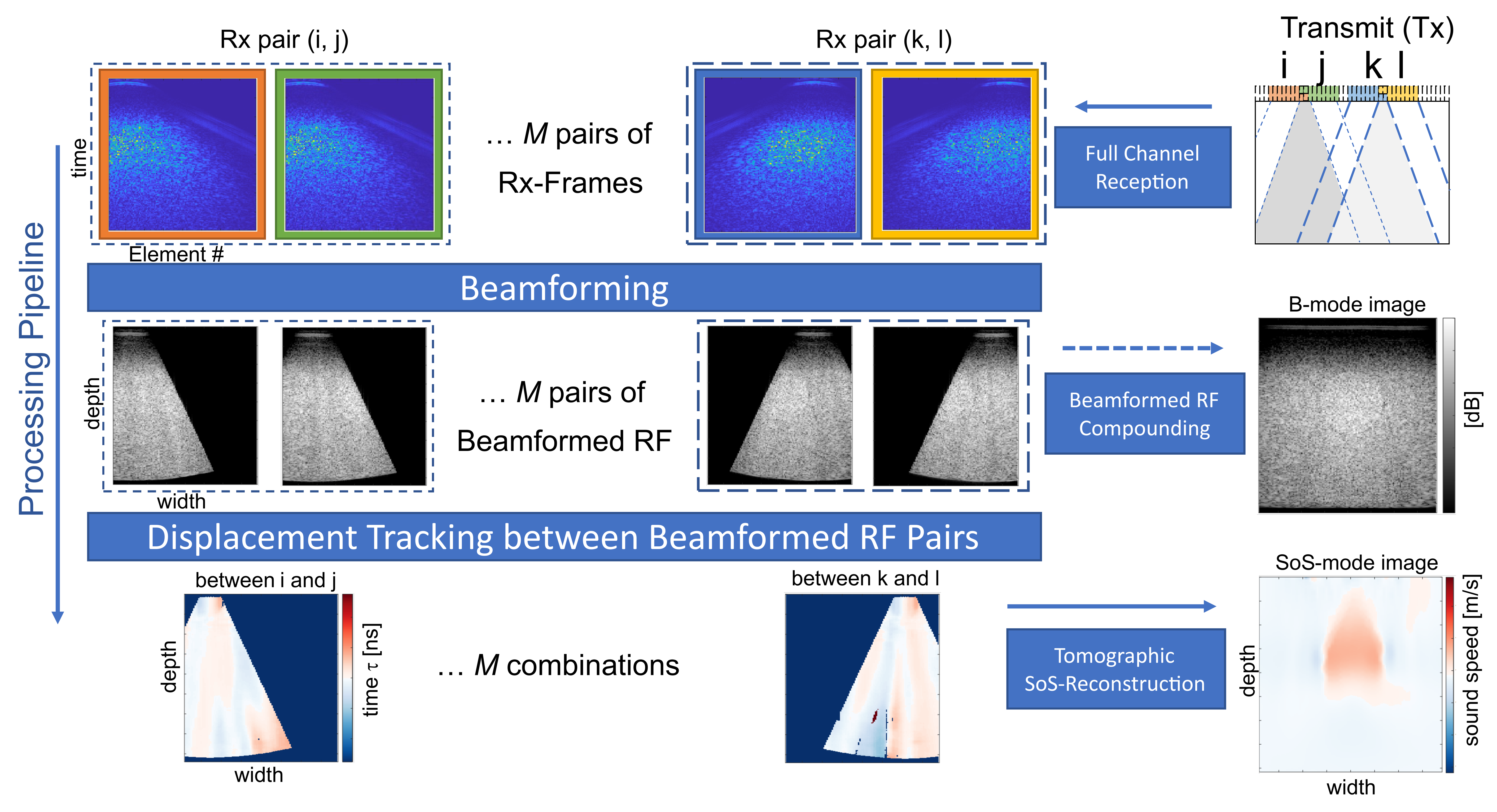}
\captionof{figure}{Processing pipeline for SoS imaging: 
Raw channel data (\textit{Rx pair}) from pairs of Transmit sequences (top-row) are beamformed using synthetic focusing (middle-row), with apparent displacements computed between each pair (bottom-row).
From these, an SoS image is reconstructed based on the forward problem of relative delays formulated given respective Tx-Rx wave paths.
For reference, we also create a B-mode image by compounding all beamformed frames.
}
\label{fig:pipelineVS}
\end{figure*}

We herein utilize the general processing and SoS reconstruction pipeline from~\cite{Rau_divergingWave_2021}.
For sake of completeness, we provide the following brief description, with an overview of processing pipeline illustrated in \Cref{fig:pipelineVS}.
In each Tx-Rx cycle the tissue is insonified by a DW (of origin SE or VS) with a given aperture $a$. 
Received echoes are then recorded for each available Rx channel as radio-frequency (RF) time data.
In case of WH, the received Rx signals are first decoded~\cite{villaverde_ultrasonic_2016} to a single-element RF data, for which SE based processing is applied thereafter.

A synthetic aperture Tx-Rx beamforming process then generates a beamformed RF frame on a fixed Cartesian spatial grid of $N_x\times N_z$, assuming a spatially-constant beamforming SoS $c_0$ and full dynamic-receive aperture (with F=1.0).

Apparent displacements are tracked between a pair of beamformed (BF) RF-frames resulting from two Tx events.
Herein we utilize a normalized cross-correlation based displacement tracking algorithm in the axial direction. 
For each image pair, this results in displacement readings as a column vector $\Delta d$ of length $N_x$$\times$$N_z$, which we map back to time-domain as $\Delta\tau=\Delta d / c_0$\,. 
Displacement readings that are outside the aperture or that are in the near field close to the transducer or that fall below a tracking noise level (assessed by a correlation coefficient threshold) are omitted from further processing. 
To increase robustness, $M$ such image pairs are used in reconstruction, with their displacements stacked as $\boldsymbol{\Delta\tau}=[\Delta\tau_1^T \cdots \Delta\tau_M^T]^T$.

A local slowness map $\boldsymbol{\hat\sigma}\in\mathbb{R}^{N_x'N_z'}$ (inverse of SoS, i.e.\ $\boldsymbol{\hat\sigma}=1/\boldsymbol{\hat c}$) is then reconstructed on a $N_x'\times N_z'$ spatial grid, via the following inverse problem~\cite{Rau_divergingWave_2021}:
\begin{equation}\label{eq:sosrecon}
 \boldsymbol{\hat\sigma} = \arg \min_{\boldsymbol{\sigma}}
 \| \textbf{L}(\boldsymbol{\sigma-}\sigma_0) - \boldsymbol{\Delta\tau} \|_1  +  \lambda \|\textbf{D}\boldsymbol{\sigma} \|_1\ \ 
\end{equation}
where the differential path matrix $\textbf{L}$$\in$$\mathbb{R}^{M N_x N_z\times N_x' N_z'}$ links the relative slowness distribution $(\boldsymbol{\sigma}-\sigma_0)$ to relative delay measurements $\boldsymbol{\Delta\tau}$, where $\sigma_0=1/c_0$ used to beamform the input RF images. 
Regularization matrix \textbf{D} together with weight $\lambda$ controls the amount of spatial smoothness and is essential due to the poor conditioning of the problem. 
For the given limited-angle computed tomographic nature, regularization \textbf{D} implements anisotropic image filtering to suppress streaking artifacts, with the gradient weighting scheme described in~\cite{sanabria_spatial_2018,Rau_divergingWave_2021}. 
The optimization problem is solved using a limited-memory Broyden–Fletcher–Goldfarb–Shanno (L-BFGS) algorithm.

\subsection{Walsh-Hadamard coded transmission for SoS imaging}

Single-element (SE) transmission can be a powerful tool, for example from a full-array of SE transmissions, known as multi-static (MST) acquisition, many transmit schemes can be synthetically simulated with time-delays and linear combinations.
Accordingly, SE diverging wave transmission has been used not only for SoS imaging, but also in other applications, e.g.\ for synthetic aperture imaging~\cite{jensen_synthetic_2006}. 
The main problem with SE transmission is the limited energy output of a single transducer element resulting in the echo amplitude reducing below the noise level within a short distance into the tissue.
To remedy such SNR issue, several Tx pulse coding schemes have been introduced, among which the Walsh-Hadamard coding was reported in~\cite{villaverde_ultrasonic_2016} to improve SNR by up to 19\,dB and it can be realized within a standard ultrasound system with a reasonable effort and sequencing access.

WH-coded transmission uses multiple TX events, at each of which all transducer elements are activated with the same signal amplitude but with positive or negative (inverted) pulse patterns based on Hadamard encoding. 
While repeating this for all Hadamard coded Tx events (equivalent to the number of elements, since this is essentially a unitary transform of MST), echo data for all Rx channels are recorded and stored.
Assuming linearity of the operated acoustic regime, the inverse Hadamard transform, i.e.\ a simple matrix multiplication, can then convert this stored data to MST equivalent format, but with much higher SNR.
This is thanks to the fact that WH-coded Tx events use many elements and possible the entire array, which deposits a larger energy into tissue compared to SE.
However, as in any synthetic refocusing process, the above decoding process assumes that all echoes originate from the same tissue locations for all Tx events.
Therefore, any motion of the transducer or the tissue during WH acquisition may cause blurring or other deterioration in the decoded Rx data, potentially negatively affecting also further processing. 

\subsection{Virtual-source transmission for SoS imaging}
Virtual source transmission has been used in synthetic aperture imaging to improve SNR and resolution in the azimuth and elevation planes~\cite{nikolov_virtual_source_2002}. 

\begin{figure}
\begin{tabular}{cc}
\includegraphics[scale=0.1]{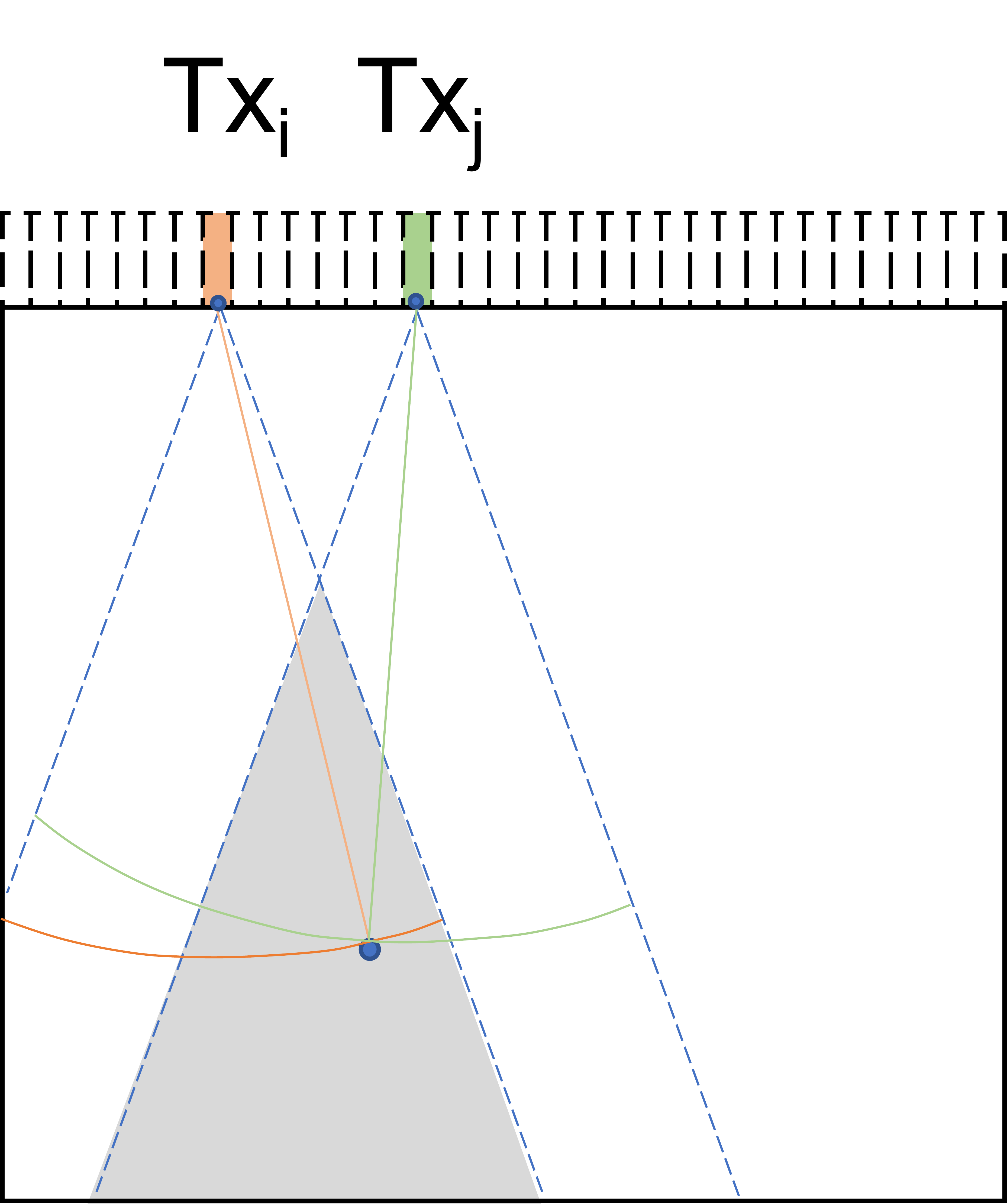} & 
\includegraphics[scale=0.1]{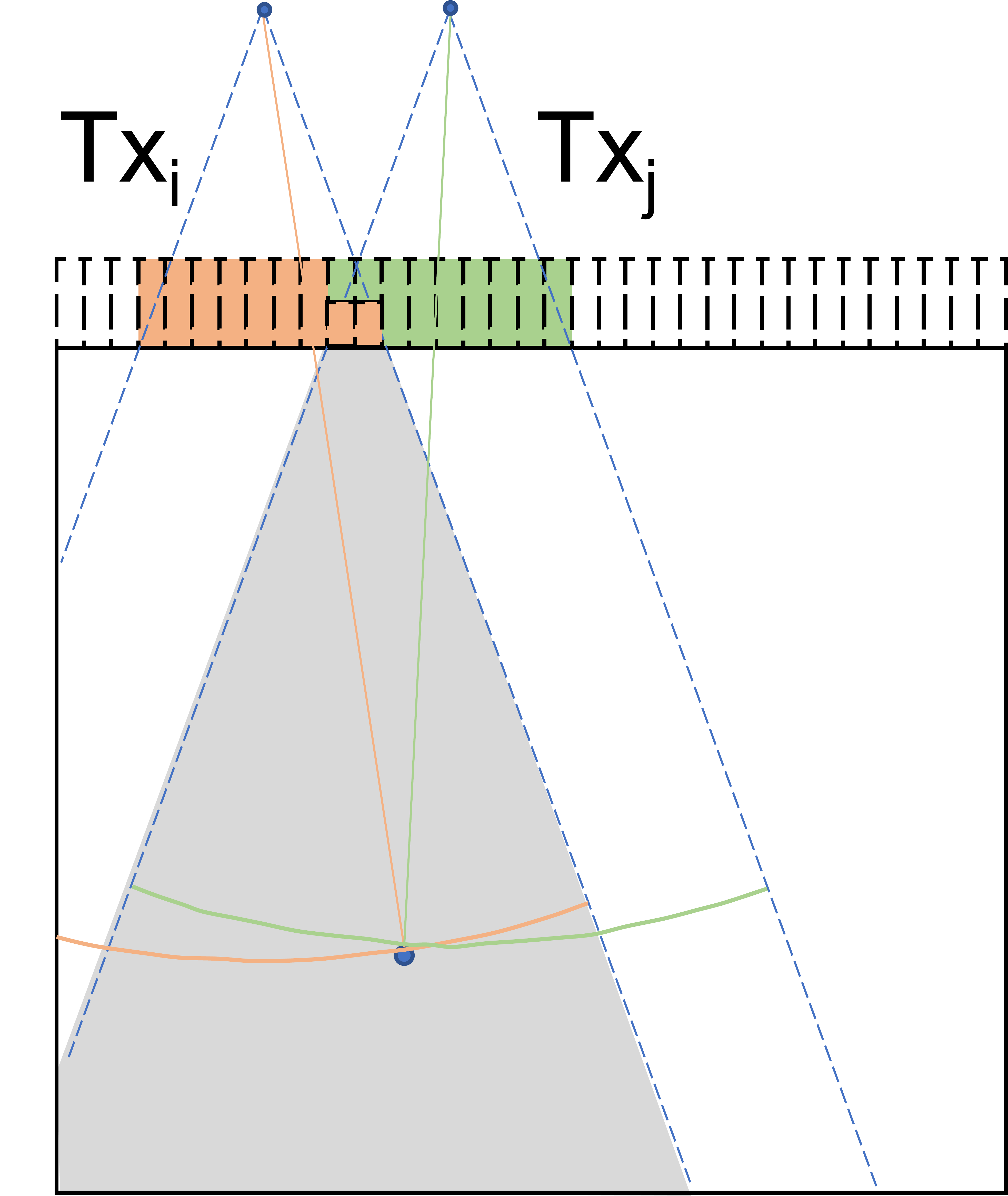} \\
(a) SE Transmit &
(b) VS Transmit
\end{tabular}
\centering
\includegraphics[scale=0.1]{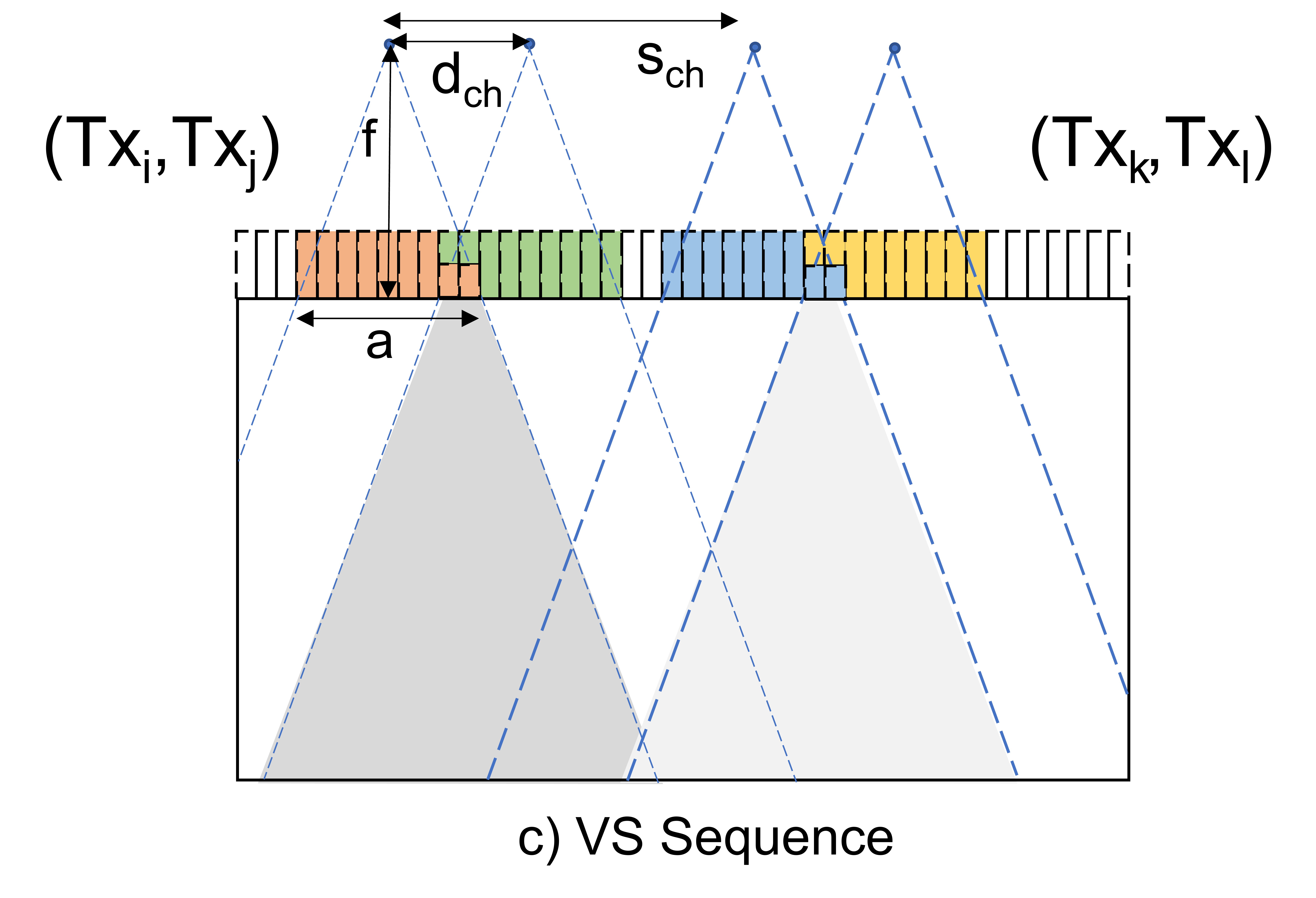}\\
(c) VS Transmit Sequences
\caption{A pair of transmits Tx$_i$ and Tx$_j$, separated by $d_\mathrm{ch}$ insonify the tissue.  (a)~For SE or decoded WH, the diverging wave Tx originates at a single transducer element on the transducer surface. (b)~For VS, the virtual focus point is behind the transducer at a distance $f$, which results in an active aperture of $a$. (c)~Consecutive Tx pairs are shifted by $s_\mathrm{ch}$ to cover the whole region of interest.
}
\label{fig:SE-VSprinciple}
\end{figure}
Multi-element VS transmission mimics the far-field of a focused transmission, where the focus point lies virtually behind the transducer surface, as seen in \Cref{fig:SE-VSprinciple}.
Active transducer element aperture $a$ is then defined by the focus point location together with the selected Tx f-number.
Element-wise Tx delays can be calculated identically to focused beam transmission, based on an assumed tissue speed-of-sound depending on the tissue under investigation.
The resulting wave front containing the energy of all transmitting elements of aperture $a$ is then a circular diverging wave.
Indeed, SE can be seen as a special case of VS with $a$=1 element and the focus point centered at that element.
Theoretical SNR gain from VS compared to SE is $20\log(\sqrt{a})$, indicating an SNR gain of 15\,dB for a 32 element VS transmission.

\subsection{Parametrizing a virtual-source sequence for SoS imaging}
For a successful VS sequence for SoS imaging, aside from obvious parameters, such as Tx center frequency, bandwidth, apodization, etc, there are a few additional implementation choices related to VS in particular. 
VS being a form of diverging wave, these implementation choices are indeed similar to those required in~\cite{Rau_divergingWave_2021}, as described below.

The first choice is the separation distance $d_\mathrm{ch}$ of the transmission pairs between which the displacements are to be computed. 
This pair separation mainly affects the disparity observed between the paired images.
The larger $d_\mathrm{ch}$ is, the more the tissues that the paired Tx passes through will differ, as a better differential input for the tomographic reconstruction.
However, with more such disparity, the displacement tracking will start to fail as the echo speckles will not correlate anymore.
In the other direction, the smaller the $d_\mathrm{ch}$, the better the displacement tracking will be, but the resulting displacements will get smaller in magnitude, which will make the solution of the inverse problem less stable; indeed approaching to null space if one were to use the same origin for both transmits.
A second effect of pair separation $d_\mathrm{ch}$ is the field of view (shaded areas in \Cref{fig:SE-VSprinciple}(c)), where the insonifications of the pair spatially overlap and thus displacement tracking can be performed.
Note that the smaller the $d_\mathrm{ch}$ is, the larger the (shaded) areas from which displacement tracking readings can be used for reconstruction, but in return -- as also explained above -- the smaller the disparity of each reading will be, and hence reducing their usability in reconstructions.
To that effect, one can see in \Cref{fig:SE-VSprinciple}(a-b) that a VS Tx pair already has a much larger (shaded) overlap area with usable displacement readings than an SE Tx pair.

Since the displacements from a single Tx pair is only sensitive to SoS variations from a small part of the entire image, more than one Tx pair is needed for imaging larger regions.
Several Tx pairs, with potentially overlapping areas, also increase the total number of measurements, potentially yielding more robust reconstructions.
Therewith, a second parameter choice becomes the number of Tx pairs $M$ to be acquired for reconstruction.
Let $s_\mathrm{ch}$ be the distance between each Tx pair, assuming without loss of generality that they are uniformly separated.
If the goal is to cover a certain imaging region of interest (ROI), which can be the entire transducer width, then $s_\mathrm{ch}$ and $M$ will be inversely proportional, i.e.\ a smaller pair separation $s_\mathrm{ch}$ would require many more pairs $M$ to cover the desired ROI.

Since our goal is to avoid or minimize negative effects from motion, we wish to minimize the total number of Tx events, while covering a large ROI with several (potentially overlapping) displacement measurement regions.
To that end, a promising parameter choice is $s_\mathrm{ch}$$=$$d_\mathrm{ch}$, i.e.\ using the second Tx of a pair as the first Tx of the next pair. 
That way, one can, for instance, perform only $M$$+$$1$ transmits to use the $M$ consecutive pairs for $M$ separate displacement measurements, which is an almost two time reduction in acquisition time from $2M$ transmits that would be needed otherwise.

\subsection{Iterative adaptation of beamforming SoS}\label{sec:iterativeSoS}
SoS reconstructions with Eq.~\eqref{eq:sosrecon} are based on displacements observed between RF images that are beamformed with a global (spatially-constant) SoS value $c_0$.
The assumption is that the local SoS values are distributed around this global value, so the beamformed RF images contain coherent speckles between which apparent displacements can be tracked.
The global SoS assumption can be set from the literature values, based on the organ, etc.
However, the SoS may change within the same anatomy, as well as across subjects~\cite{sanabria_speed_2018_sarc}, with pathology~\cite{Ruby_breast_19}, or even during the lifetime of the same subject (e.g., menstrual cycle for the breast~\cite{Ruby_breast-density_19}).
If the global assumption for beamforming is far from the average SoS in imaged tissue, the beamformed image quality may degrade and therewith the tracked displacements and hence the reconstructed SoS images.

\begin{figure}
  \centering
    \includegraphics[width=0.6\linewidth]{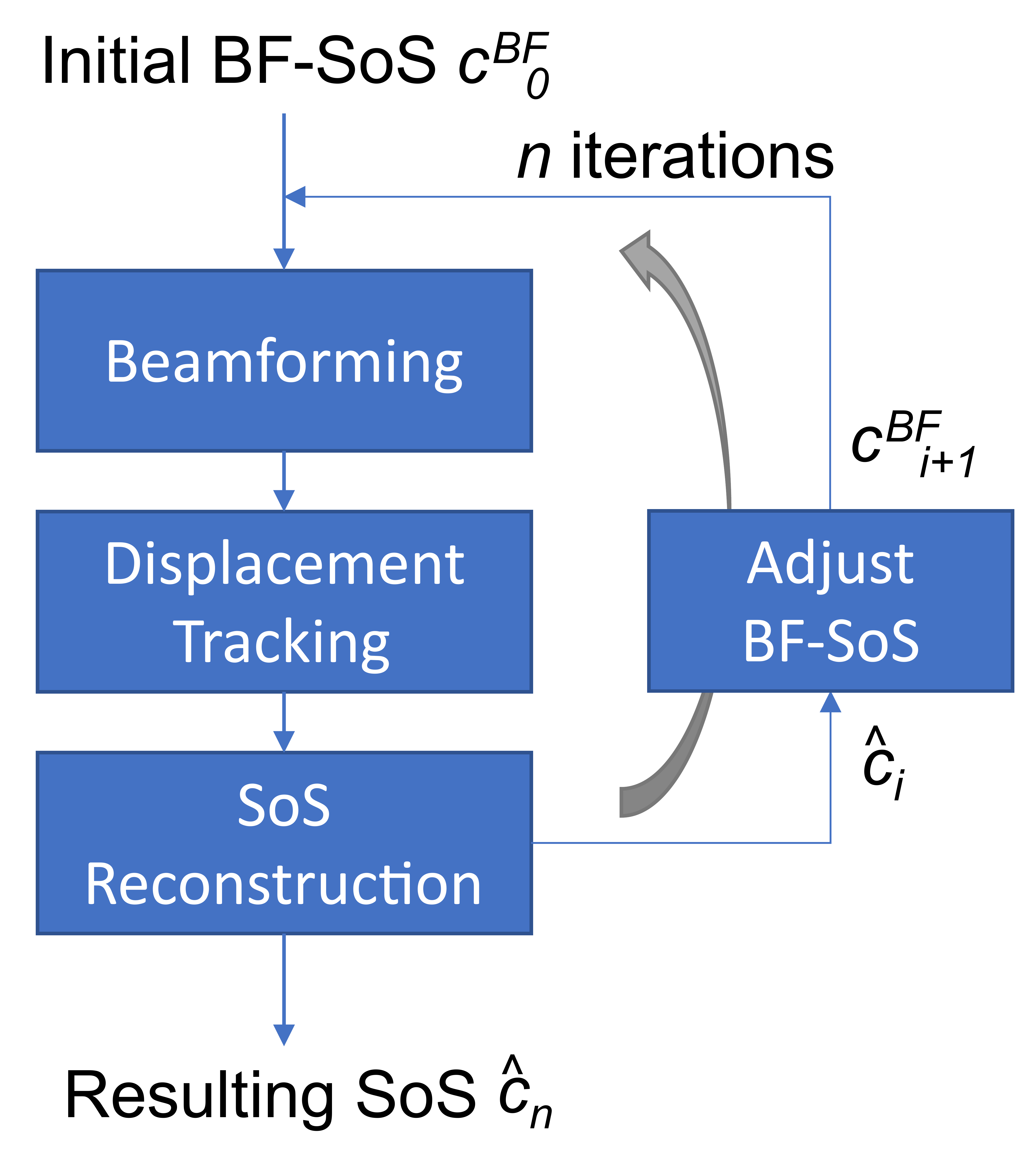}
\captionof{figure}{Iterative SoS Reconstruction: 
The beam-form SoS is updated after each iteration until it converges towards the tissue SoS.
}
\label{fig:IterativeSoS}
\end{figure}

We propose to perform SoS reconstructions iteratively, by readjusting the beamforming SoS based on the current reconstructed values, as shown in \ref{fig:IterativeSoS}.
Our hypothesis is that with an incorrect beamforming SoS, even if the local SoS values may be inaccurate, this will provide us a better overall SoS to improve the beamforming.
To minimize large deviations from individual SoS results, we use statistically robust operations for such beamforming SoS update:
In particular, for a reconstructed local SoS image $\boldsymbol{\hat{c}_i}$ of resolution $N_x'\times N_z'$, we use the \emph{median} value $\hat{c}_i$ of this SoS image to update the beamforming SoS for iteration $i$ as a smoothed time average as follows:
\begin{equation}
    c^\mathrm{BF}_{i+1} = \frac{(c^\mathrm{BF}_i + \hat{c}_i)}{2} \hspace{10pt} i= 1 \dots n\ \ .
\end{equation}
This process is initialized with $c^\mathrm{BF}_1=c_0$ for the first iteration, and repeated for $n$ iterations, with the local SoS reconstruction from the last iteration taken as the resulting output image.

\section{Experiments}
Numerical simulations, tissue-phantom experiments, and in-vivo data acquisition have been conducted to evaluate our proposed methods.
In particular, we first demonstrate the negative effect of motion to motivate our solutions, then we compare SE, WH, and VS sequences, while we also show the value of setting beamforming SoS iteratively.

\subsection{Data acquisition system}
\label{sec:RADAsystem}
Experiments were conducted using an UF-760AG ultrasound system (Fukuda Denshi, Tokyo, Japan) with 64 Rx channels.
We used a FUT-LA385-12P linear array transducer with $N_c = 128$ elements and 300\,$\mu$m pitch.
For each Tx, a 4 half-cycles pulse of $f_c=5$\,MHz center frequency is transmitted, followed by a reception of RF data. 
The received data is stored temporarily in element-wise buffers during the acquisition time from the deepest imaged location, and then transported over a high-speed data-link to an attached PC for storage and data-processing, before the next Tx-Rx cycle is performed.
This buffer transport leads to a period 37.5\,ms between two consecutive Tx events.

\subsection{Numerical simulations}
\begin{figure}
  \centering
    \includegraphics[width=\linewidth]{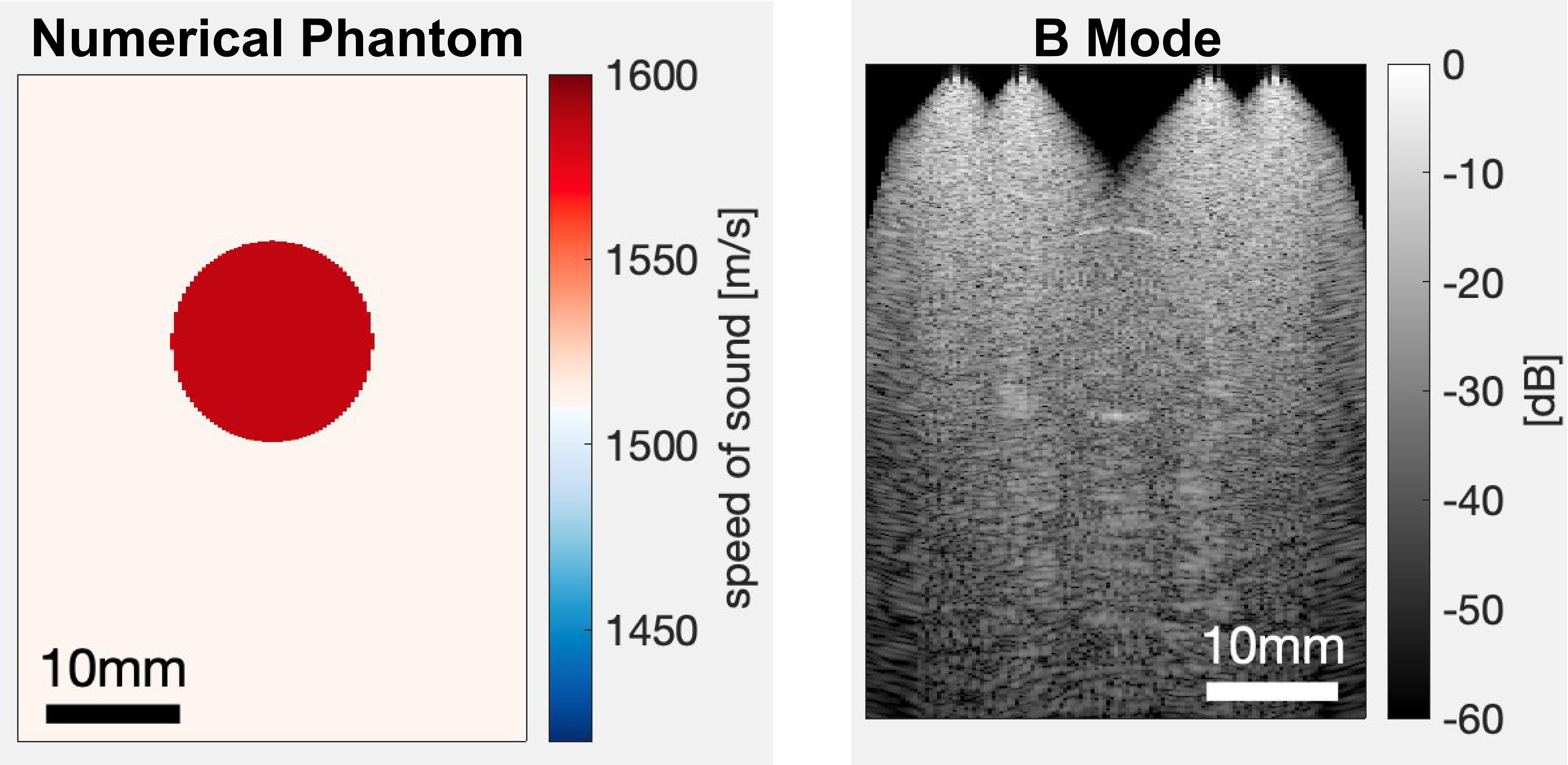}
\captionof{figure}{Numerical simulation setup: (a)~the numerical phantom with an inclusion and (b)~the B-mode image generated from this by simulating in k-Wave two SE Tx pairs: (24,\,41) and (88,\,105).
}
\label{fig:K-WavePhantom}
\end{figure}

Numerical simulations were performed in Matlab with k-Wave ultrasound toolbox~\cite{treeby_k-wave:_2010}.
Tissue medium was discretized on a grid with 25\,$\mu$m spatial resolution.
A circular inclusion of 1585\,m/s and 15\,mm diameter was modeled on a background substrate of 1510\,m/s (i.e.\ 5\% contrast), as seen in \Cref{fig:K-WavePhantom}.
To generate speckle, minor density variations were added in the entire domain, similarly to~\cite{Rau_divergingWave_2021}.
In simulations, we model the acquisition settings as well as the transducer (FUT-LA385-12P, Fukuda Denshi, Tokyo, Japan) used in the physical experiments.

With numerical simulations, our goal was to demonstrate quality degradation even in a controlled and noise-free synthetic setting, where full-matrix Rx reception can be conducted.
This helps to motivate the need for motion-insensitive acquisition schemes, regardless of physical setup limitations.
Since noise, hence SNR, then becomes no issue in simulations, we herein only study the SE sequence, and only using a small number of Tx pairs with $M=2$ as shown in \Cref{fig:K-WavePhantom}.

After each Tx cycle, a full-matrix Rx is performed, and then the phantom is moved laterally or axially by $t$ number of simulation grid pixels, before performing the next Tx-Rx cycle. 
Larger motion speeds are simulated by moving the grid by larger amounts.
To isolate the effect of motion in this simulated environment, we applied motion either only between different Tx pairs (i.e.\ the scene assumed static between the two transmits of each pair) or only within the Tx pairs.
Note that this ignores any motion occurring during the time of flight of a Tx pulse within the tissue.
This is an acceptable assumption, since this happens multitudes times faster than the pair acquisition, thus yielding negligible motion effect.
Furthermore, such motion during Tx time-of-flight already happens in other imaging modalities, such as in elastography, without major detrimental effect on motion estimation.

\subsection{Tissue-mimicking phantom experiments}

\begin{figure}
  \centering
    \includegraphics[width=\linewidth]{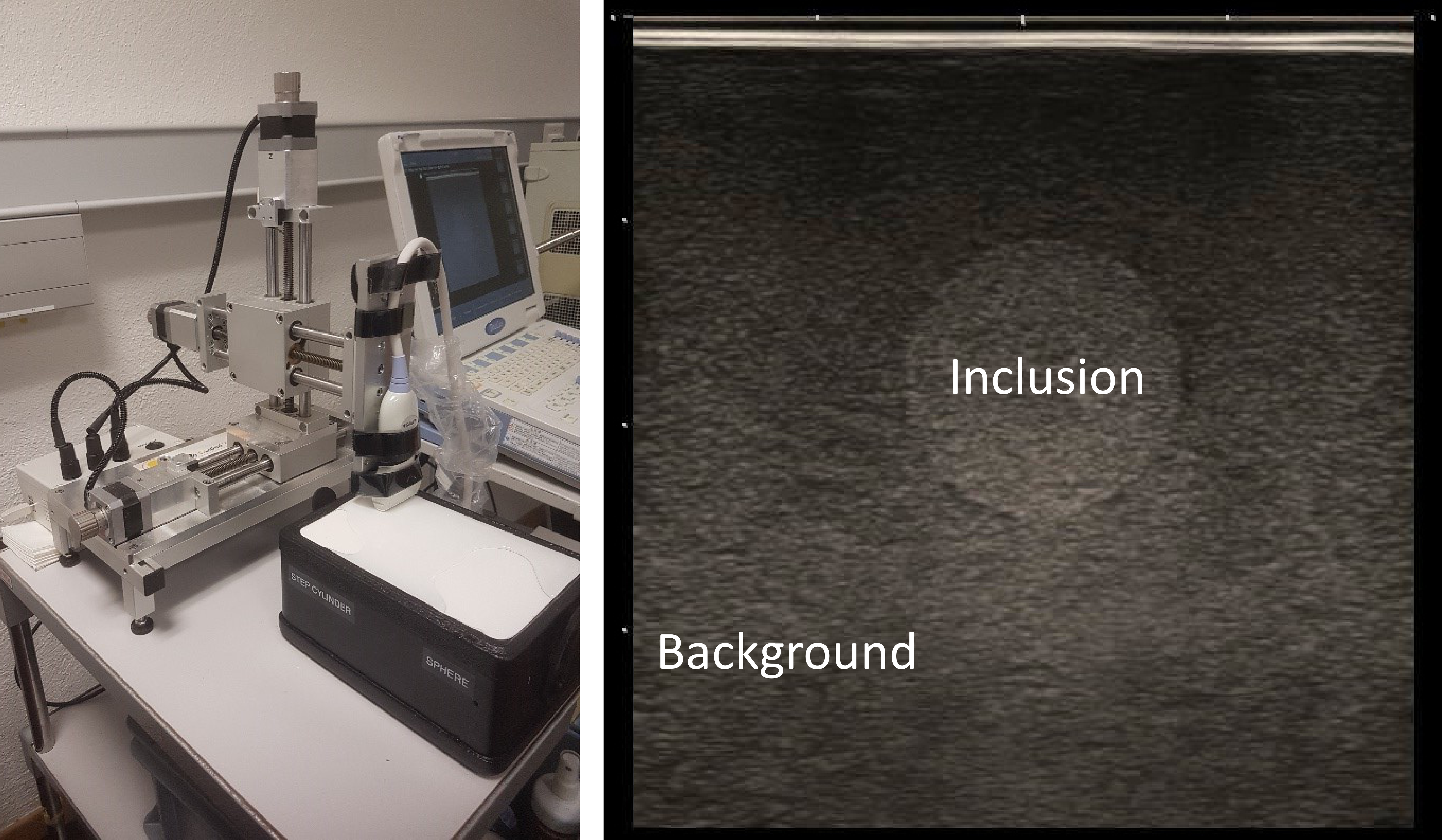}
\captionof{figure}{Controlled linear motion stage and SoS-Tissue-mimicking Phantom
}
\label{fig:LinStagePhantom}
\end{figure}

For this study, a custom SoS phantom with is built by CIRS (Norfolk, VA, USA) with the following specifications:
Within a background substrate of 1515\,m/s, a cylindrical inclusion of 1585\,m/s (4.6\% contrast) is embedded with a diameter of 10.3\,mm centered at a depth of 15\,mm from the surface.
For controlled linear motion, the transducer was fixed to a three-axis motion stage as shown in \Cref{fig:LinStagePhantom}(left).
In order not to deform the phantom during motion, a water layer of $\approx$2\,mm was applied on the phantom surface below the transducer. 
Translation experiments were conducted separately in lateral (x-axis) or axial (z-axis) directions, with translation speeds up to 1\,mm/s which the setup allowed.
We synchronized the Tx-Rx sequences to start right after the motion start, used these acquired data frames during linear motion in our reconstruction, and then reset the transducer position to its original location before the next motion experiment.

Motion of the transducer relative to the phantom induces a spatial shift between the beamformed RF frames within a Tx pair that is used for displacement estimation. 
Such shift of the scene depends upon the time between two Tx events. 
In a commercial ultrasound system implementation which allows for the transport of all acquired channel data to RF frame buffer in real-time during reception, such time then depends only on the reception depth. 
In our custom raw-data acquisition system, however, such transport needs 32 times longer than such ideal case. 
Therefore, a speed of 1\,mm/s in our experimental setup can be interpreted in-effect equivalent to 32\,mm/s motion considering an ideal ultrasound system implementation with sufficient transport capability.

\subsection{In-vivo experiments}
In-vivo data from breast lesions was collected in a clinical study at Kantonsspital Baden, Switzerland, under ethics approval, external monitoring, and informed patient consent.  
During data collection, the operator first used the B-mode images for probe navigation to a frame where a suspicious lesion was visible, and started the SoS acquisition sequence using a foot pedal while keeping the transducer as steady as possible. 
Data from SE, WH, and VS sequences were acquired automatically one after the other. 

\subsection{Evaluation Metrics}\label{sec:EvalMetrics}
For a quantitative analysis of the SoS reconstruction in numerical and phantom experiments, we used root mean squared error RMSE=$\sqrt{\frac{1}{N}\sum(\hat{\boldsymbol{c}}-\boldsymbol{c}^\star)^2}$\,, where $\boldsymbol{c}^\star$\, is the groundtruth SoS map. 
In the phantom experiments, the inclusion visible in the B-mode image was delineated, to set the groundtruth SoS values inside and outside based on the manufacturer-reported values.

For evaluating in-vivo experiments, due to lack of ground-truth SoS values, we used contrast
$\Delta$SoS=$|\mu_\mathrm{inc} - \mu_\mathrm{bkg}|$\,, i.e.\ the absolute difference of median $\mu$ of the inclusion SoS and the median $\mu$  of the background SoS.
We used median to be statistically robust to potential superfluous image pixel readings.
The inclusions were annotated on B-mode by a clinician and the background is considered as the locations outside the inclusion plus a margin of 5\,mm to omit border inaccuracies in inclusion delineations and reconstructions.

\subsection{Implementation settings}
For all physical experiments with the phantom and in-vivo, the following parametrizations have been used:
\subsubsection{SE Sequence}
This implies an aperture of $a=1$.
To be able to study different parameter combinations retrospectively, during physical acquisitions we collected all element combinations, i.e.\ a multi-static (MST) acquisition with the number of Tx events $n_\mathrm{TX}$=128. 
For the presented reconstructions, we used Tx pairs with $d_\mathrm{ch}$=17 and $s_\mathrm{ch}$=10, resulting in $M$=12 pairs as \{(1,18), (11,28), \ldots, (111,128)\} aligned with the findings of~\cite{Rau_divergingWave_2021}.
\subsubsection{WH Sequence}
This is implemented following~\cite{villaverde_ultrasonic_2016}. 
Since our transducer has twice as many elements than the 64 Tx/Rx channels available on our system, to collect the data for a complete 128-element Hadamard coding, we had to use $n_\mathrm{TX}$=4x128 Tx events, for the combinations of Tx and Rx for the left and right transducer sub-apertures separately. 
By Hadamard decoding this WH data to its MST equivalent, we could thereafter use the above procedures for SE sequence reconstruction.
Accordingly, the above Tx pair parameters for SE reconstruction could also be used here. 
\subsubsection{VS Sequence}
\begin{figure}
  \centering
    \includegraphics[width=0.8\linewidth]{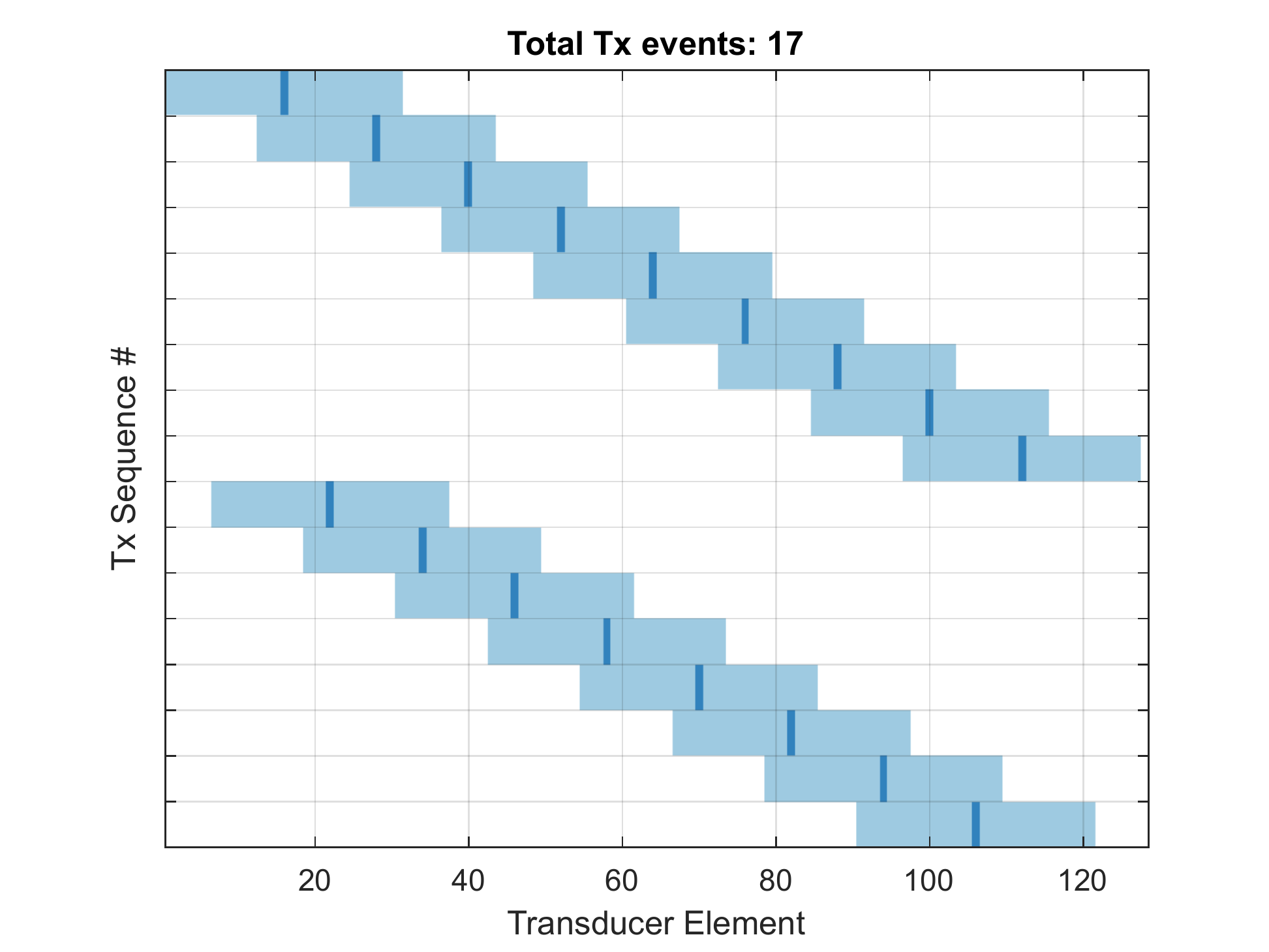}
\captionof{figure}{A fast VS transmit sequence, with pair separation $d_\mathrm{ch}$=$s_\mathrm{ch}$=12, yielding $n_\mathrm{TX}$=9 Tx events and $M$=15 given two passes over the transducer surface with the second pass centered between the first. For each Tx event (row), the light blue area indicates the Tx aperture $a$ with the darker blue marker indicating the lateral position of the VS center.
}
\label{fig:Tx_Sequence}
\end{figure}

With a diverging wave Tx focus depth $f$=-0.9\,mm and Tx f-number=-1, an effective Tx aperture of $a$=31 elements was used.
After preliminary testing on phantoms, we chose a pair separation $d_\mathrm{ch}$=$s_\mathrm{ch}$=12 given the trade-off between sufficient disparity and good displacement tracking between the pairs. 
With these, we could cover the 128 element transducer surface with $n_\mathrm{TX}$=9 Tx events, yielding $M$=8 pairs.
Given the very short time this fast acquisition takes, we were able to easily increase the number of collected pairs, which we did using a second pass over the transducer surface.
We offset the second pass locations between the original pass, as illustrated in \Cref{fig:Tx_Sequence}, in order to collect more diverse data for reconstruction pairs.
As seen in the figure, this leads to $n_{TX}=17$  Tx events leading to $M=15$ fame pairs, using consecutive transmits except between the 8th and 9th Tx events which are far apart.

\subsubsection{SoS Reconstruction}
For a fair comparison of sequences, we adopted the following common choice of parameters to process data from all sequences:
Full RX aperture was used for RF beamforming; near-field within the first 5\,mm was masked out from displacement estimations; and the regularization weight $\lambda$ was set to $0.065$\,. 

\section{Results and Discussion}
\subsection{Iterative adaptation of beamforming SoS}
\begin{figure*}
\begin{subfigure} {\textwidth}
\centering
\includegraphics[width=0.9\linewidth]{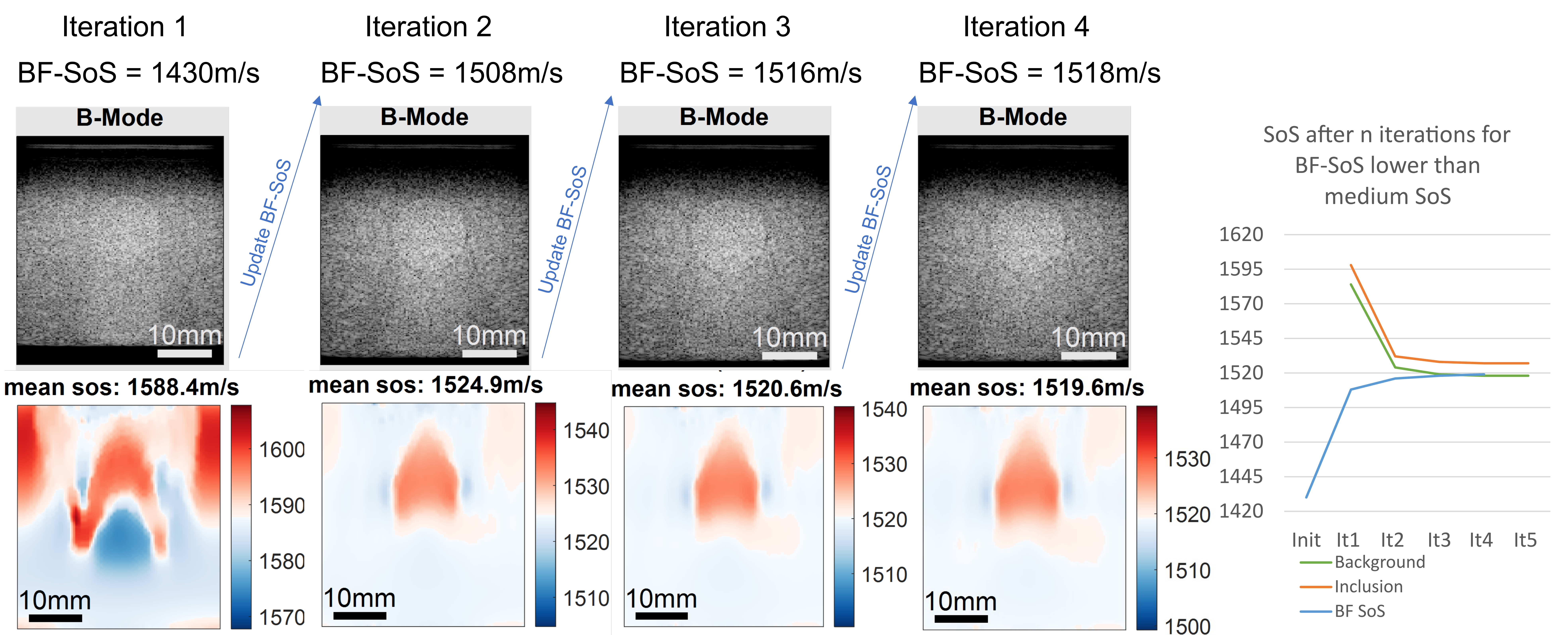} 
\caption{Beamforming SoS initialized lower than the known medium SoS}
\label{fig:subim1_IerativeSoS}
\end{subfigure}
\begin{subfigure} {\textwidth}
\centering
\includegraphics[width=0.9\linewidth]{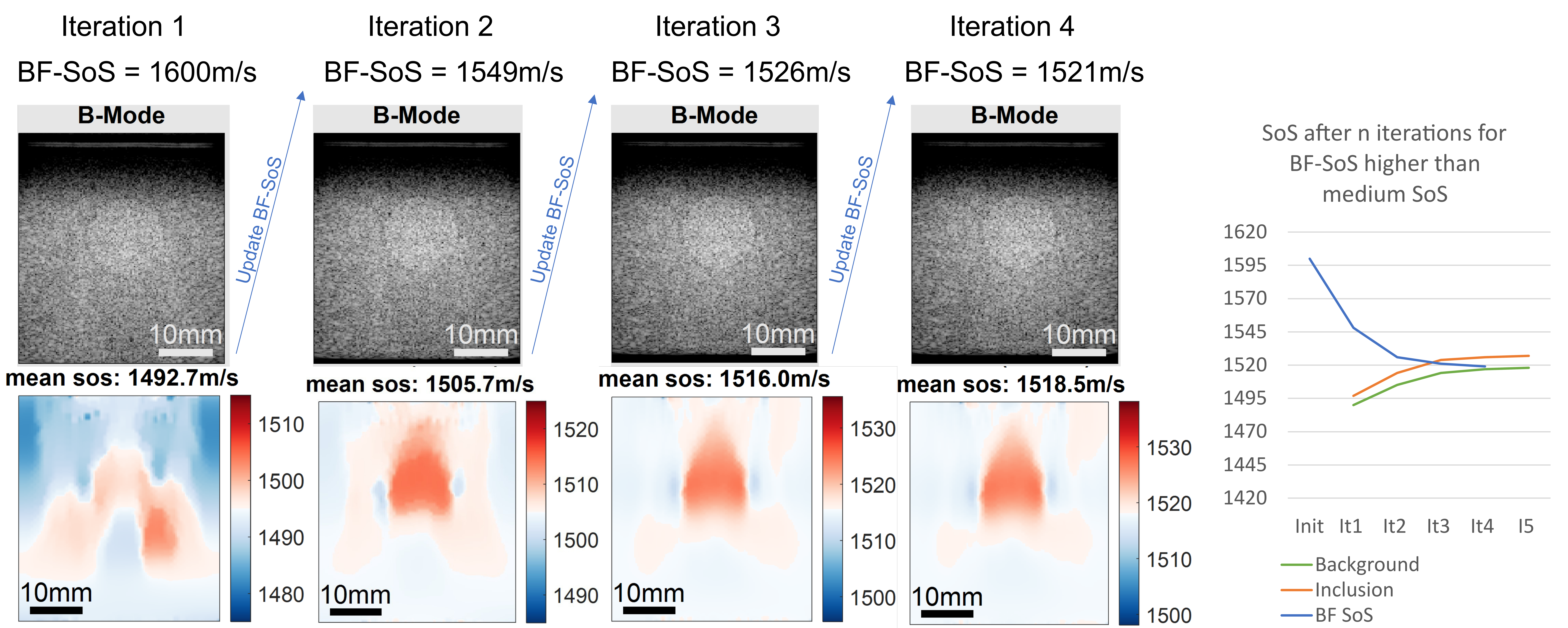} 
\caption{Beamforming SoS initialized higher than the known medium SoS}
\label{fig:subim2_IerativeSoS}
\end{subfigure}
\captionof{figure}{Reconstructions and beamforming (BF-)SoS evolution during iterative adaptation of BF-SoS, initialized with assumed values (a)~lower and (b)~higher than the known medium SoS.
Plots on the right show BF-SoS progression (blue), together with the mean SoS reconstruction values inside (red) and outside (green) the inclusion.
}
\label{fig:SoSAdaptation}
\end{figure*}

To study the effect of this proposed method, we acquired data with the transducer fixed in place.
\Cref{fig:SoSAdaptation} shows the beamforming (BF-)SoS and the resulting image reconstructions over iterations, starting from low and high initial BF-SoS assumptions.
The reconstructions are seen to improve over the iterations, with the BF-SoS and the reconstructed values converging over time.
In all further results, we used $n$=3 iterations, which is seen here to approximate the background SoS within $0.5\%$ of the manufacturer-declared value of 1515\,m/s.
We used this SoS known value to initialize BF-SoS in all further phantom experiments, so the sequences can be evaluated hereafter rather than the effect of incorrect BF-SoS assumptions. 

\subsection{Freehand imaging experiments}\label{sec:FreehandArtefacts}
Before our controlled experiments with a motion-stage, we study the artifacts from freehand imaging of the tissue-mimicking phantom.
Five data-sets were collected for each sequence, with:
\begin{figure*}
\centering
\includegraphics[width=0.8\linewidth]{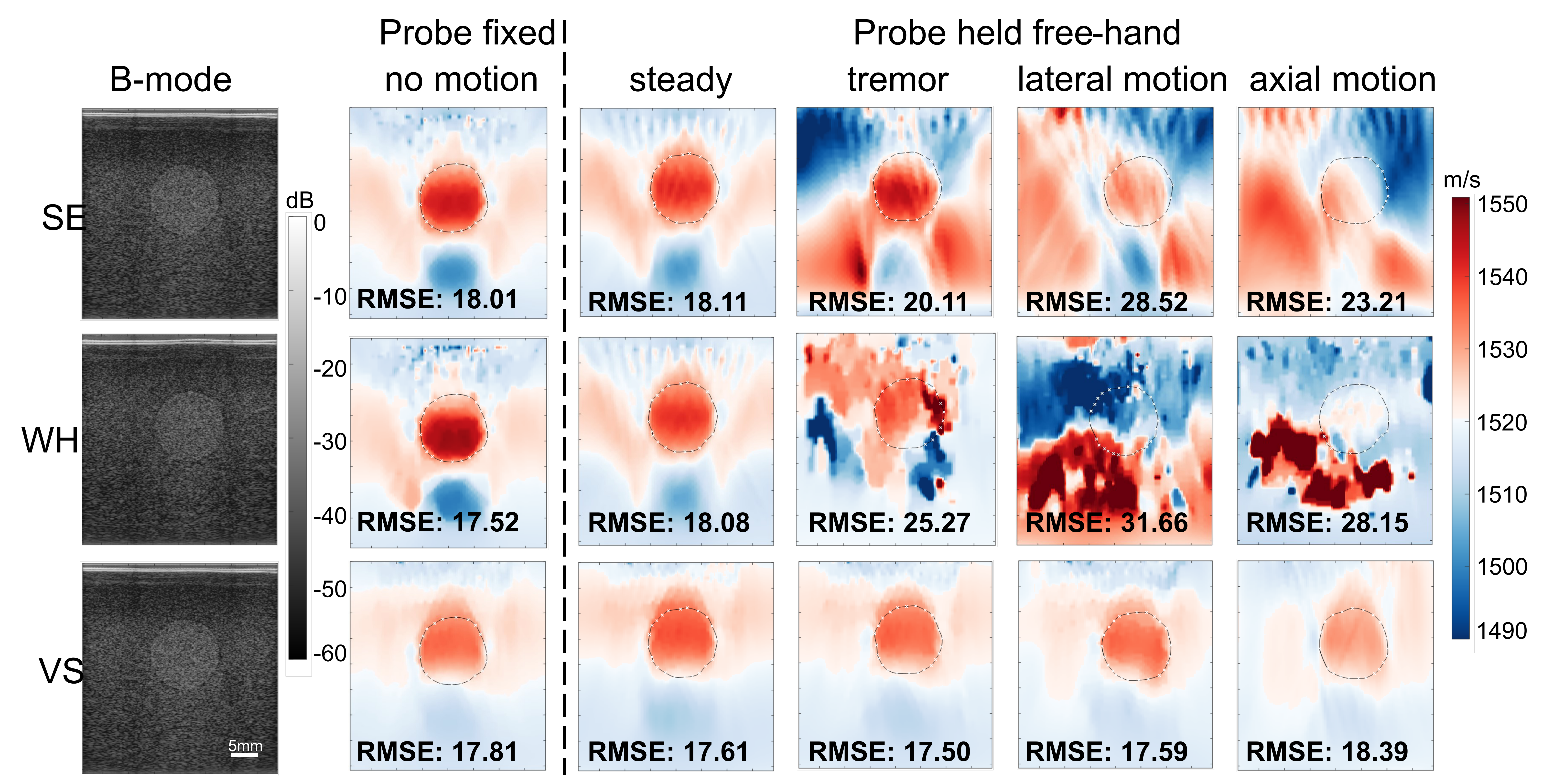} 
\captionof{figure}{Motion artefacts under free-hand conditions
}
\label{fig:MotionFreehand}
\end{figure*}
    (1)~the probe fixed mechanically in place;
    (2)~the probe held by the operator as still as possible by grounding the forearm;
    (3)~the probe held relaxed by the operator and moved gently in place, causing minor hand tremors;
    (4)~the probe moved by the operator laterally at a small pace, as in an ultrasound exam; and
    (5)~the probe moved by the operator axially at a small pace, as in gentle compression during an exam.
Each data acquisition was after centering the probe over the same SoS inclusion seen in \Cref{fig:MotionFreehand}.
From the presented results, it is seen that the larger the motion is, the more difficult it is for the SoS reconstruction. 
WH sequences is seen to deteriorate the most, since the Hadamard decoding of each SE equivalent Tx event requires the linear combination of all received WH RF frames, which become most inconsistent over the motion.
VS is seen to be the most robust to motion.

\subsection{Controlled imaging experiments}
To quantify the effect of motion, we conducted the following controlled experiments. 

\subsubsection{Motion Simulation}
For simulating motion, we moved the numerical phantom by \{0,1,2,3\} simulation grid pixels in lateral and axial directions.
We separately simulated the motion between each Tx event, i.e.\ both within and between Tx pairs, as well as only the motion between the pairs.
The latter is to study the effects of motion between reconstruction pairs collected further in time.
\begin{figure*}
\centering
\includegraphics[width=0.8\linewidth]{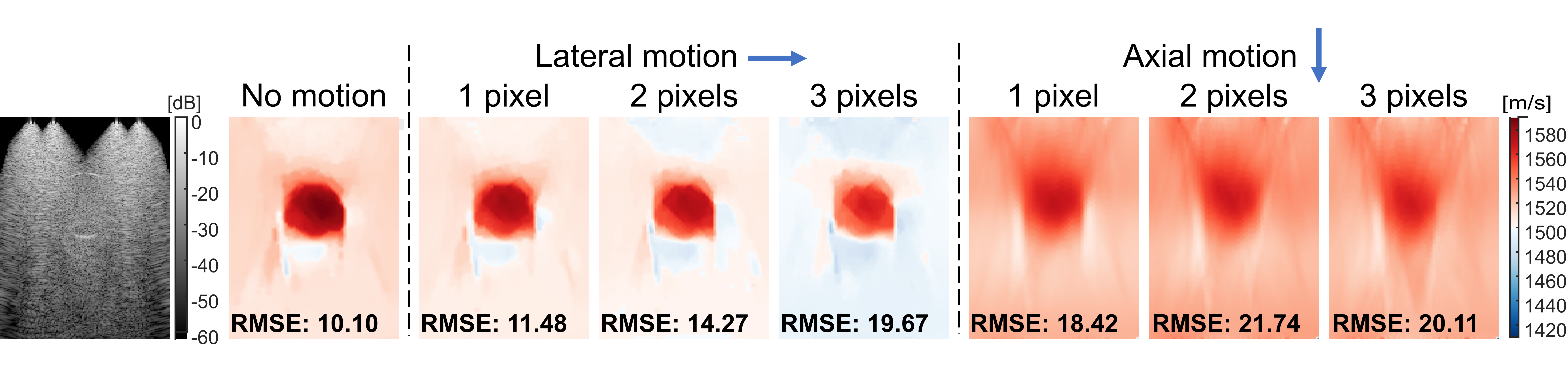} 
\captionof{figure}{Numerical simulation results showing artefacts with increasing motion for 25$\mu m$ per pixel.}
\label{fig:MotionSimulated}
\end{figure*}
From the results tabulated in \Cref{tabresultsSim}, it is seen that such motion happening only between the pair acquisition has little to no effect in reconstruction accuracy, whereas given within-pair motion as well presents a major detrimental effect, also demonstrated with reconstructions in \Cref{fig:MotionSimulated}.
This can be explained by the fact that the motion within a pair confounds displacement tracking, impeding reconstructions; whereas motion between the pairs merely affects the spatial alignment/overlap of separate pairs solved within the same linear system, which is not of a major concern given the reconstruction resolution, regularization, etc.
Note that for our simulation grid of 25\,$\mu$m/pixel, tabulated within-pair motions simulate the speeds \{0,\,613,\,1226,\,1839\}\,$\mu$m/s, respectively, on a physically-equivalent setup with our settings for the multi-static acquisition of the SE sequence.
In both \Cref{tabresultsSim} and \Cref{fig:MotionSimulated}, the reconstruction quality is seen to degrade with increasing motion, and the axial motion is seen to have a larger influence compared to the lateral motion.

\begin{table}[b]
\begin{center}
\caption{RMSE results [m/s] from simulated motion, applied only between Tx pairs or applied between each Tx event (i.e., a typical continuous motion both within and between Tx pairs).}
\label{tabresultsSim}
\renewcommand{\arraystretch}{1.2}
\begin{tabular}{|l|l|c|c|c|c|}
\hline
\multirow{2}{*}{\parbox[c]{14ex}{\textbf{Simulated\\ motion}}} & \multirow{2}{*}{\textbf{Direction}} & \multicolumn{4}{c|}{\textbf{Motion [pixels]}} \\ \cline{3-6}
& & \textbf{0} & \textbf{1} & \textbf{2} & \textbf{3}  \\
\hline \hline
\multirow{2}{*}{\parbox[c]{14ex}{Only between\\ Tx pairs}} & Lateral & \multirow{4}{*}{$10.10$} & $10.44$ & $10.17$ & $10.10$ \\
\cline{2-2} \cline{4-6}
& Axial & & $10.23$ & $10.11$ & $10.15$  \\
\cline{1-2} \cline{4-6}
\noalign{\vskip\doublerulesep\vskip-\arrayrulewidth}
\cline{1-2} \cline{4-6}
\multirow{2}{*}{\parbox[c]{14ex}{Between each\\ Tx event}} & Lateral &  & $11.48$ & $14.27$ & $19.67$ \\
\cline{2-2} \cline{4-6}
& Axial &  & $18.42$ & $21.74$ & $20.11$  \\
\hline
\end{tabular}
\end{center}
\end{table}
\subsubsection{Controlled Motion with Linear Stage}

\begin{figure*}
\centering
\includegraphics[width=0.8\linewidth]{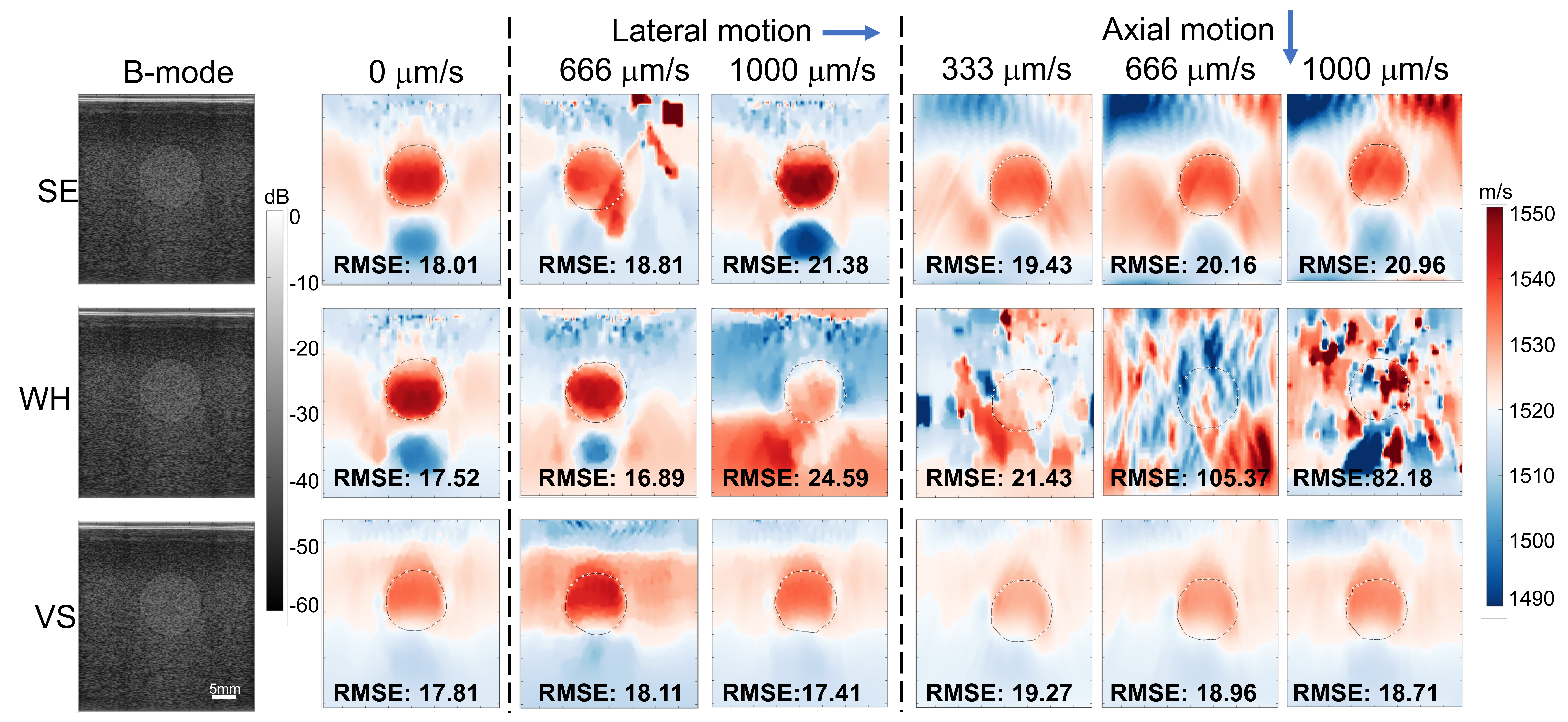} 
\captionof{figure}{Phantom experiments with controlled linear motion in lateral and axial directions with different speeds.}
\label{fig:MotionControlled}
\end{figure*}

Results from controlled lateral and axial transducer motion are given in \Cref{fig:MotionControlled} demonstrate increased degradation with larger motion speed. 
Among the sequences, WH sequence results degrade the most with motion, concordant with observations from free-hand experiments.
Consistent with the simulation findings, axial motion is seen here to be of larger detriment compared to lateral motion.
This is likely due to the fact that any misalignments between Tx pairs caused by motion can then confound the misalignments due to SoS differences, where the latter are encoded in the Tx direction, thus mainly presented in the axial direction in which we also conduct the displacement tracking.

\begin{table}
\begin{center}
\caption{RSME results [m/s] from controlled motion stage.}
\label{tabresultsLinMotion}
\renewcommand{\arraystretch}{1.2}
\begin{tabular}{|l|l|c|c|c|c|}
\hline
\multirow{2}{*}{\textbf{Direction}} & \multirow{2}{*}{\parbox[c]{8ex}{\textbf{TX\\ sequence}}} & \multicolumn{4}{c|}{\textbf{Motion speed [$\boldsymbol\mu$m/s]}} \\ \cline{3-6}
& & \textbf{0} & \textbf{333} & \textbf{666} & \textbf{1000} \\
\hline \hline
\multirow{3}{*}{Lateral} & SE & $18.01$ & \multirow{3}{*}{---} & $18.81$ & $21.38$  \\
\cline{2-3}
\cline{5-6}
             & WH & $17.52$ &  & $16.89$ & $24.59$ \\
\cline{2-3}
\cline{5-6}
             & VS & $17.81$ &  & $18.11$ & $17.41$  \\
\hline \hline
\multirow{3}{*}{Axial} & SE & $18.01$ & $19.43$ & $20.16$ & $20.96$  \\
\cline{2-6} 
             & WH & $17.52$ & $21.43$ & $105.37$ & $82.18$ \\
\cline{2-6} 
             & VS & $17.81$ & $19.27$ & $18.96$ & $18.71$  \\
\hline
\end{tabular}
\end{center}
\end{table}

\subsubsection{In-vivo SoS Imaging}
For initializing BF-SoS in the in-vivo experiments, due to a lack of a good initial estimate, we used the following strategy:
For each patient, an additional acquisition with the three sequences was carried out at a different breast location that is relatively homogeneous without any visible lesions.
For these, we reconstructed images separately with data from each sequence for $n$=3 iterations, each initialized by an approximate breast SoS value of 1450\,m/s. 
BF-SoS from the last iteration (i.e.\ the median of the last reconstruction) is then considered as a good estimate of the patient-specific overall breast background SoS -- known to change per patient, age, menstrual cycle, etc~\cite{sanabria_breast-density_2018,Ruby_breast-density_19}.
To increase robustness, we repeated the above for two different breast background locations, and used their average BF-SoS to initialize the iterative reconstruction process for the inclusion view with the corresponding sequence.
Running this inclusion specific reconstructions also for $n$=3 iterations, so that the BF-SoS for the inclusion view is further optimized, we arrived at the in-vivo spatial SoS reconstruction of that inclusion view for a given sequence.

\begin{figure}
\centering
\includegraphics[width=\linewidth]{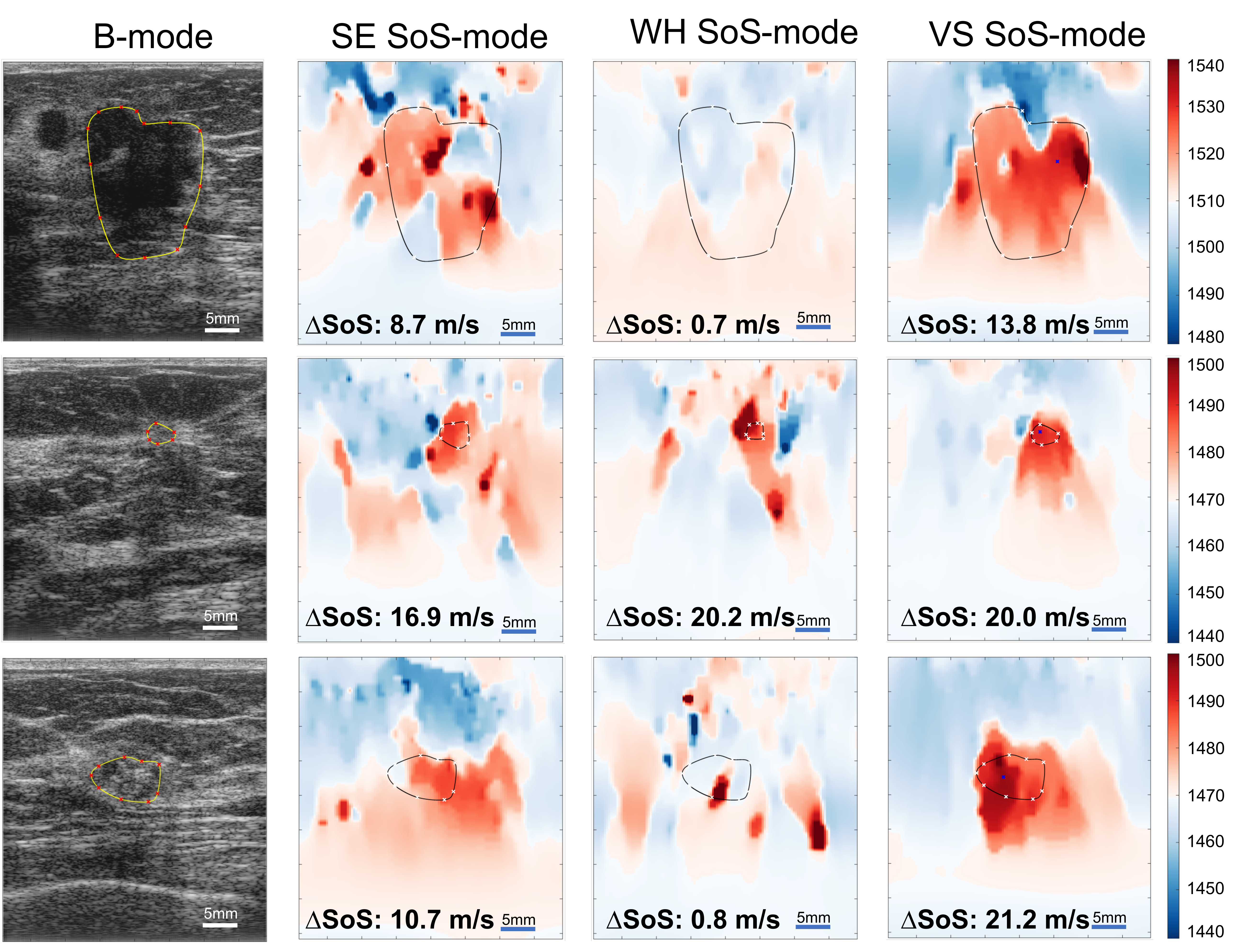} %
\captionof{figure}{SoS reconstruction results from in-vivo data.}
\label{fig:MotionInVivo}
\end{figure}

\Cref{fig:MotionInVivo} shows the reconstructions from three lesions of biopsy-confirmed ductal carcinoma with different sizes and locations.
For the sequences SE, WH, and VS, the average contrast $\Delta$SoS across these three inclusions are 12.1, 7.2 and 18.3\,m/s, respectively; demonstrating highest contrast with our proposed fast VS sequence.

\section{Conclusion}
Ultrasound imaging using a hand-held transducer requires robustness against motion, caused by the operator or by inherent physiological sources such as breathing and heart-beat. 
We propose herein virtual-source transmits for SoS image reconstruction.
This uses transmit pulses that can emit much higher energy compared to single-element transmission, thus providing high SNR raw RF data, while obviating a need for decoding encoded sequences, as in Walsh-Hadamard sequence, thereby minimizing the motion confounding the displacement readings.
We herein present a specific sequence parametrization and implementation based on virtual-source transmits, which further minimizes the total time of acquisition while providing many number of Tx pairs for robust reconstructions.
With our results, we present the proposed sequence to be superior to the compared alternatives regarding motion robustness necessary for in-vivo applications.
Note that improved local SoS reconstructions can also help better correct aberrations in beamforming as shown in~\cite{rau_ultrasound_2019}.

Beamforming SoS may largely affect the local SoS reconstruction results, as also demonstrated in~\cite{Bezek_global_22}.
Accordingly, we herein employ an iterative approach for finetuning the beamforming SoS, using the median reconstruction SoS for beamforming in the next iteration. 
We present this to successfully identify the beamforming SoS in a phantom study.
Note that such an iterative method can be used in a real time clinical application for consecutively acquired frames, thereby reducing additional acquisition and computation time, as the overall image content and hence the image-specific beamforming SoS can be expected to change relatively slowly with respect to the frame rate, with the probe manipulation during an ultrasound exam.  
We will next study the imaging technique proposed herein and its clinical diagnostic value in a large clinical cohort.

\section{Acknowledgments}
Research funding was provided by the Swiss National Science Foundation.
Bayer Schweiz AG and BRACCO Suisse SA provided funding for clinical study administration; and partial funding provided by the Young Researchers Grant of the European Society of Breast Imaging.
The authors would like to thank for the help and support by the Kantonsspital Baden personnel, in particular Dr.\ Monika Farkas, Dr.\ Anna Potempa, and the study nurse Silke Callies.

\bibliographystyle{IEEEtran}
\bibliography{arxivRefs}

\end{document}